# Robust topological edge states at the perfect surface step edge of topological insulator ZrTe$_5$

Xiang-Bing Li[1†], Wen-Kai Huang[1†], Yang-Yang Lv[2†], Kai-Wen Zhang[1], Chao-Long Yang[1], Bin-Bin Zhang[2], Y. B. Chen[1*], Shu-Hua Yao[2], Jian Zhou[2], Ming-Hui Lu[2], Li Sheng[1,3], Shao-Chun Li[1,3*], Jin-Feng Jia[3,4], Qi-Kun Xue[5], Yan-Feng Chen[2,3], Dingyu Xing[1,3]

[1] *National Laboratory of Solid State Microstructures, School of Physics, Nanjing University, Nanjing, Jiangsu 210093, China*

[2] *National Laboratory of Solid State Microstructures, Department of Materials Science and Engineering, Nanjing University, Nanjing, Jiangsu 210093, China*

[3] *Collaborative Innovation Center of Advanced Microstructures, Nanjing University, Nanjing 210093, China*

[4] *Key Laboratory of Artificial Structures and Quantum Control (Ministry of Education), Department of Physics and Astronomy, Shanghai Jiao Tong University, Shanghai 200240, China*

[5] *State Key Laboratory of Low-Dimensional Quantum Physics, Department of Physics, Tsinghua University, Beijing 100084, China*

†These authors contributed equally to this work.

*Corresponding Authors: ybchen@nju.du.cn; scli@nju.edu.cn.

**We report an atomic-scale characterization of ZrTe$_5$ by using scanning tunneling microscopy. We observe a bulk bandgap of ~80 meV with topological edge states at the step edge, and thus demonstrate ZrTe$_5$ is a two dimensional topological insulator. It is also found that an applied magnetic field induces energetic splitting and spatial separation of the topological edge states, which can be attributed to a strong link between the topological edge states and bulk topology. The perfect surface steps and relatively large bandgap make ZrTe$_5$ be a potential candidate for future fundamental studies and device applications.**

Two-dimensional topological insulator (2D TI) is a new type of quantum matter, and hosts quantum spin Hall effect (QSHE) [1,2]. It is featured by an insulating bulk band gap and time-reversal-invariant topological edge states (TESs) protected against localization and backscattering. Since the discovery of QSHE in the inverted HgTe/CdTe quantum wells [3,4], progress has been made in predictions and characterizations of 2D TI materials [5-21]. Most of the currently confirmed 2D TIs is of limited practical use, due to either small bulk band gaps or the difficulty in achieving thin sheets down to single layer. Searching for a 2D TIs material with practically ideal properties is still of great importance, which, however, is rather challenging. Furthermore, the response of topological edge states (TESs) to an applied magnetic field has not been fully understood. Theoretical models have been established to address the spatial distribution of TESs under magnetic fields [22,23], but the energetic evolution has never been studied. Tuning the external magnetic fields or gating voltages, the transition from a 2D TI to a normal insulator or a quantum Hall state could be realized [4,21].

Recently, Weng *et al* [24] predicted that single layer $ZrTe_5$ is a 2D TI with a large band gap, and the 3D crystal of $ZrTe_5$ is located near the phase boundary between weak and strong topological insulators, sensitive to lattice parameters. In contrast, several experimental studies suggested that $ZrTe_5$ might be a Dirac semimetal [25-27]. Therefore, a direct characterization is rather important to clarify the bulk band topology of $ZrTe_5$. In fact, studies of $ZrTe_5$ can be tracked back to two decades ago, due to its anomalous magnetoresistance and low-temperature thermoelectric power [28-30].

In this study, we use scanning tunneling microscopy (STM) to characterize the surface of cleaved single crystal $ZrTe_5$. We observe not only a gapped bulk band structure at the surface terrace of *a-c* plane, but also TESs located at the step edge. It then follows that the $ZrTe_5$ surface is a 2D TI, and that the $ZrTe_5$ crystal is a weak 3D TI consisting of stacked layers of 2D TIs, in good agreement with the theoretical prediction[24]. In the presence of magnetic field, the edge states undergo a large energy splitting, which could be explained by time-reversal-symmetry-broken edge states close linked to the bulk band topology [22,23,31].

Single crystals $ZrTe_5$ were grown by the chemical vapor transport method with iodine ($I_2$) as the transport agent. Polycrystalline $ZrTe_5$ was first synthesized by a solid-state reaction (at about 500 °C for 7 days) in a fused silica tube sealed under vacuum (~ 4×10$^{-6}$ mbar). A ratio of 1:5 for the high purity Zr (99.999%) and Te powder (99.999%) was adopted. The mixture of prepared ZrTe$_5$ polycrystalline and I2 (~ 5 mg/L) powder were loaded into a sealed evacuated quartz tube, and then put into a two-zone furnace. After growing in a temperature profile of 520-450 °C for over 10 days, the single crystal was successfully grown, with a typical size of ~35 × 1 × 0.5 mm$^3$.

All STM characterizations were carried out in ultrahigh vacuum (UHV) with a Unisoku LT-STM at ~4 K. The base pressure was 5×10$^{-11}$ mbar. ZrTe$_5$ single crystal was cleaved in situ in UHV and quickly cooled down to 4K, prior to STM/STS measurement. Constant current mode was adopted. dI/dV spectra were taken using a lock-in amplifier. A modulation of 5 mV at 1000Hz was applied. A mechanically polished PtIr tip was used.

The ZrTe$_5$ crystal has an orthorhombic layered structure [32], and contains 2D sheets of ZrTe$_5$ in the *a-c* plane, which stack along the b axis via interlayer van de Waals interactions, as shown in Fig. 1a. Each 2D sheet consists of alternating prismatic ZrTe$_3$ chains along the *a* axis that are linked by parallel zigzag Te chains. The prism of ZrTe$_3$ is formed by a dimer of Te and an apical Te atom surrounding a Zr atom.

ZrTe$_5$ is easily cleavable along the *a-c* plane, while it also exhibits quasi1D preference along the *a* axis. Figures 1b and 1c show the topography of such a cleaved ZrTe$_5$ surface, in which Te dimer and Te apical atoms can be well identified. The zig-zag chain of Te atoms is right intercalated in between. Note, no charge density wave phase is observed at either 4K or 80K.The steps are dominantly along the *a* axis due to the quasi-1D preference, and they are super-straight in mesoscopic scale (~a few μm, see Fig. S1[33]). Such long and straight surface steps make the contact to other quantum material, such as superconductor or ferromagnetic metal, easily controllable. The measured height of a single step is ~0.8 nm. The present STM results are in good agreement with the crystal structure of ZrTe$_5$ [24]. Steps along the *c* axis are seldom observed, but not straight at all (see Fig. S2 [33]).

The local density of states (LDOSs, from spectroscopic measurement) obtained at surface terrace is plotted in Fig. 1d and S3 [33]. Its high homogeneity indicates the high quality of ZrTe$_5$ single crystal. An energy gap as large as ~80 meV is identified, with the top of valence band and the bottom of conductance band located at ~-35 meV and ~+45 meV, respectively. This result will be further proved below by the QPI analysis at the step edges. The Fermi level is located inside the band gap, but slightly towards the valence band, indicating that the surface terrace is a nearly intrinsic semiconductor. The spectroscopic feature at the ZrTe$_5$ terrace agrees well with the DFT calculation for single layer ZrTe$_5$ [24]. Furthermore, no observable residue states were detected in the band gap region, supporting the model of weak TI rather than strong TI. If it were a strong TI, residual intensity due to topological surface states would be detectable within the energy gap [24].

To ascertain the electronic structure at the step edge, the LDOSs along a line perpendicular to the step edge is plotted in Fig. 2a. Clearly, the quasi-particle interference, due to confinement by the step, is observed in valence and conductance band regions, indicating a 2D-like bulk band structure. In Fig. 2b, the corresponding 1D fast Fourier transform shows the scattering interference dispersion. The dispersion below the Fermi energy resembles to the recent ARUPS results [25]. An energy gap can be also identified between the dispersions for valence band and conductance band. When approaching the step edge, a couple of new discontinuous peaks appear in the gap region (see also the in-gap feature in the FFT). In contrast to the spectra at terrace, the intensity is pronouncedly enhanced near the step edge and spans the whole gap region, indicating the existence of TESs [1,2]. The penetration depth, as measured in the profile perpendicular to the step edge in Fig. 2c, is ~7.7 nm and ~6.5 nm, for energy near and away from the Fermi level respectively. This subtle difference can be understood by the argument that the wave function of edge state penetrates further into the terrace as its energy is closer to the valence or conductance band. The STS measurements at the step along the *c* axis are analyzed in supplemental materials (see Fig. S4)[33].

Figure 2d shows two STM spectra obtained in the surface terrace region (black triangle in Fig. 2a) and at the step edge (red triangle in Fig. 2a), respectively. The broad bumps located in the valence and conductance band regions arise from the interference at the step edge. The peaks in the gap region are asymmetric, and the 1D character can be further confirmed by fitting the peaks with a 1D density of states exhibiting an inverse-square-root singularity. This characteristic of the edge states recalls the theoretical calculations for two different types of cuttings [24]. According to the calculated results, observation of inverse-square-root singularities is expected for step edge terminated by the prismatic chain rather than by the zigzag chain [24]. Our spectra are more consistent with the prismatic chain terminated steps. Due to the tip curvature effect, it is difficult to identify the exact atomic geometry at the step edge.

The 1D character of the edge states can be further revealed by the spatially resolved spectroscopic mapping along another step. The edge states stay exactly along the step edge, as shown in Fig. 2e. Since the penetration depth of edge states is relatively large, the edge channels can circumvent large-sized perturbation of a few nm, as shown in Fig. 2e and S5, which is just comparable to the spatial resolution limits for micro-fabrication techniques. Moreover, the edge states can be observed at all kinds of steps (see Fig. S5)[33], unlike in the Bi bilayer where the edge states are sensitively coupled to the step geometry[19]. Therefore, the edge states can in principle form looped electron channels along the periphery of a $ZrTe_5$ sheet, which can be easily made by micro-fabrication technology, and are rather robust against external perturbations.

An applied magnetic field normal to the surface will break the time-reversal-symmetry, and drive a complicated evolution of the edge state, as shown in Fig. 3a (also Fig. S6)[33]. Three characteristic peaks, marked as A, B, and C, are tracked for increased magnetic fields. Each peak at energy E is split into two branches with energy difference ΔE, the corresponding energies shifting up and down, respectively. The intensity for the higher-energy branch is lower than that for the lower-energy one, under relatively high fields. We rule out the possibility of observed bulk Landau quantization based on the following arguments. First, the peak splitting emerges mainly at the step edge, not observable away from the step edge (see Fig. 3c). Second, the splitting is prominent in the gap region where there is no bulk band state. Finally, the evolutions of E (Fig. 3b, S7) and ΔE (Fig. 3d, S7) with applied field show neither linear nor parabolic dependence. If the evolutions are the bulk effect, both E vs B and ΔE vs B will exhibit either a linear or parabolic behavior, respectively, corresponding to the bulk Dirac or normal electrons. As a result, the STS spectra in Figs. 3b, 3d, and S7 do not come from the bulk band structure, but the TESs. For edge state $A_1$, the above argument seems somewhat lack of convincing evidence, since its energy is close to the top of valence band and there may be a coupling between edge state $A_1$ and the bulk state. However, states B and C are right inside the band gap and well isolated from the bulk. It was theoretically suggested that an applied magnetic field may open a tiny gap of the edge states at the Dirac point. However, such a tiny gap cannot be distinguished in the present experiment due to the complication of the STS spectra.

Owing to the time-reversal symmetry, the two helical edge states have the symmetric dispersion for opposite momentums, k and –k, and the identical spatial distribution. Thus the two edge states are not distinguishable by STM in the absence of magnetic field. When the time-reversal symmetry is broken by an external magnetic field, it was theoretically suggested that the two helical edge states persist, but are delocalized from the step edge. The favored edge state is pushed further toward the step edge and the un-favored one gradually merges into the bulk [22,23,31]. This scenario can qualitatively explain the decrease in intensity for the higher energy branch of two splitting peaks. Regardless of the nonlinear dependence of $\Delta E$ on B (see Fig. 3d), a fitting with $\Delta E = \bar{g}\mu_B B$ roughly gives an effective g-factor of 40 ~ 60, suggesting that the splitting arises from not only the Zeeman effect, but also the orbital effect due to vector potential [27,34]. Furthermore, the effective g-factor is comparable to the value for bulk Landau quantization obtained in a magneto-infrared spectroscopy study[27]. Therefore, we conclude that the edge state evolution is highly correlated with the bulk band topology. The schematic shown in Fig. 3e illustrates the magnetic-field-induced Landau quantization and Zeeman splitting of the bulk band, which results in an effective splitting of the corresponding edge states. This simple model can qualitatively explain the experiment, but in fact the evolution will be much more complicated if the bulk band and edge state symmetries are taken into account, as seen from the unusual dependence of $\Delta E$ on B. Further theoretical and experimental investigations, especially under higher magnetic field, are required to fully understand the physical mechanism.

In summary, we have shown that the ZrTe$_5$ surface is a 2D TI with bulk band gap and topological edge states located at the surface step edge. It then follows that the ZrTe$_5$ crystal is a weak 3D TI rather than a Dirac semimetal, which is formed by stacking the 2D topological sheets. This material is easily cleavable, the band gap of the surface terrace is relatively large, and the perfect surface step edges obtained are ultra-long and -straight. The topological edge states are found rather robust against non-magnetic perturbations. In addition, we have also studied the evolution of edge states in the presence of an increased magnetic field, which can be understood by a theoretical model of the time-reversal-symmetry-broken topological edge states.

The work was supported by the State Key Program for Basic Research of China (Grants No. 2014CB921103, No. 2015CB921203, No. 2013CB922103), National Natural Science Foundation of China (Grants No.11374140, 11374149, 51032003, 50632030, 10974083, 51002074, 10904092, and 51472112), and the New Century excellent talents in University (NCET-09-0451).

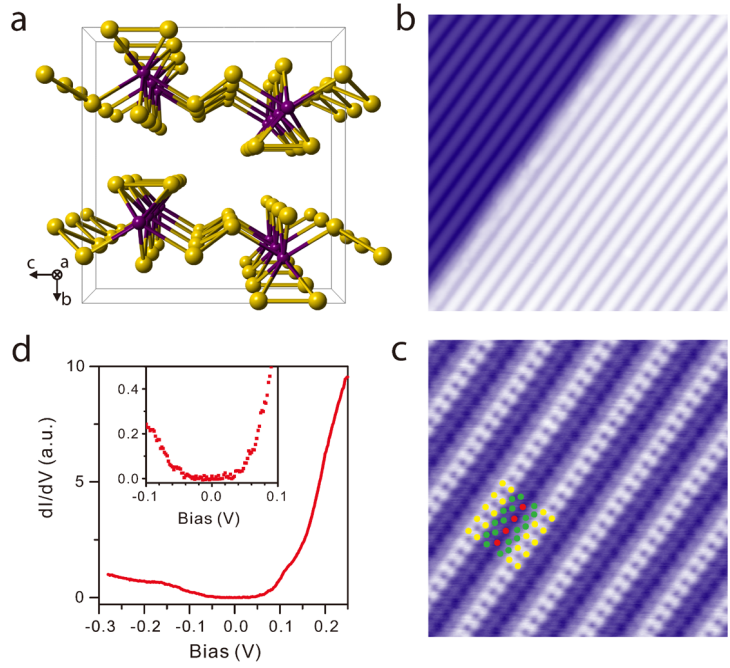

FIG. 1 (color online). (a) Crystal structure of ZrTe$_5$. (b) STM image of 25 × 25 nm$^2$ obtained at 4K (bias voltage U = +1.0 V, tunneling current I$_t$ = 100 pA). The height of the monolayer step is ~0.8 nm. (c) Atomic-resolution image of ZrTe$_5$ surface (8×8 nm$^2$; U = +350 mV, I$_t$ = 130 pA). The yellow balls mark the Te dimers, the red apical Te atoms, and the green zig-zag Te atoms. (d) Differential conductance (dI/dV spectra, U = +250 mV, I$_t$ = 200 pA, Vosc = 5 mV) taken over the terrace and far away from the step edge.

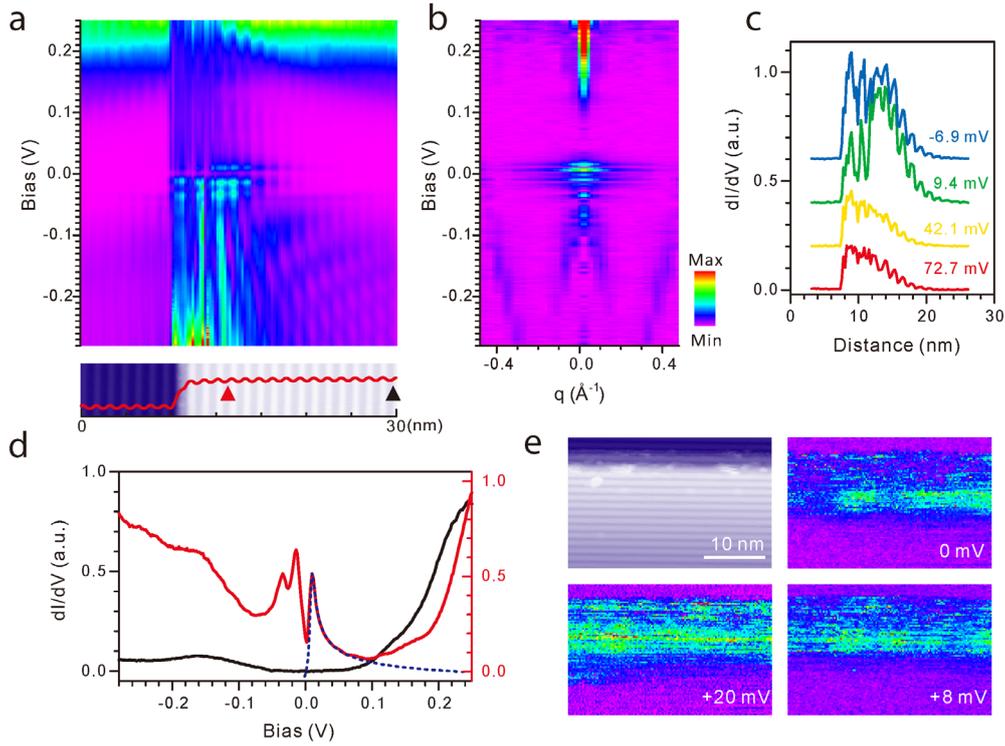

FIG. 2 (color online). (a) Differential conductance (dI/dV spectra, U = +250 mV, $I_t$ = 200 pA, $V_{osc}$ = 5 mV) taken across the step. The corresponding step topography is shown below, and the red line profile illustrates the step geometry (30nm in length). (b) 1D Fast Fourier Transform of a, showing the quasi-particle interference. (c) Line cuts at various energies extracted from a, showing the variation of penetration depth over energy. (d) Two spectra extracted from a, at the positions marked by black and red triangles. The dotted blue line is the fitting results with a 1D density of states exhibiting the inverse-square-root singularity. The broadening used is ~3 meV. (e) STM topography (34 × 23 nm$^2$; U = +1.05 V, $I_t$ = 100 pA) of another step and the corresponding dI/dV mapping taken at various bias voltages.

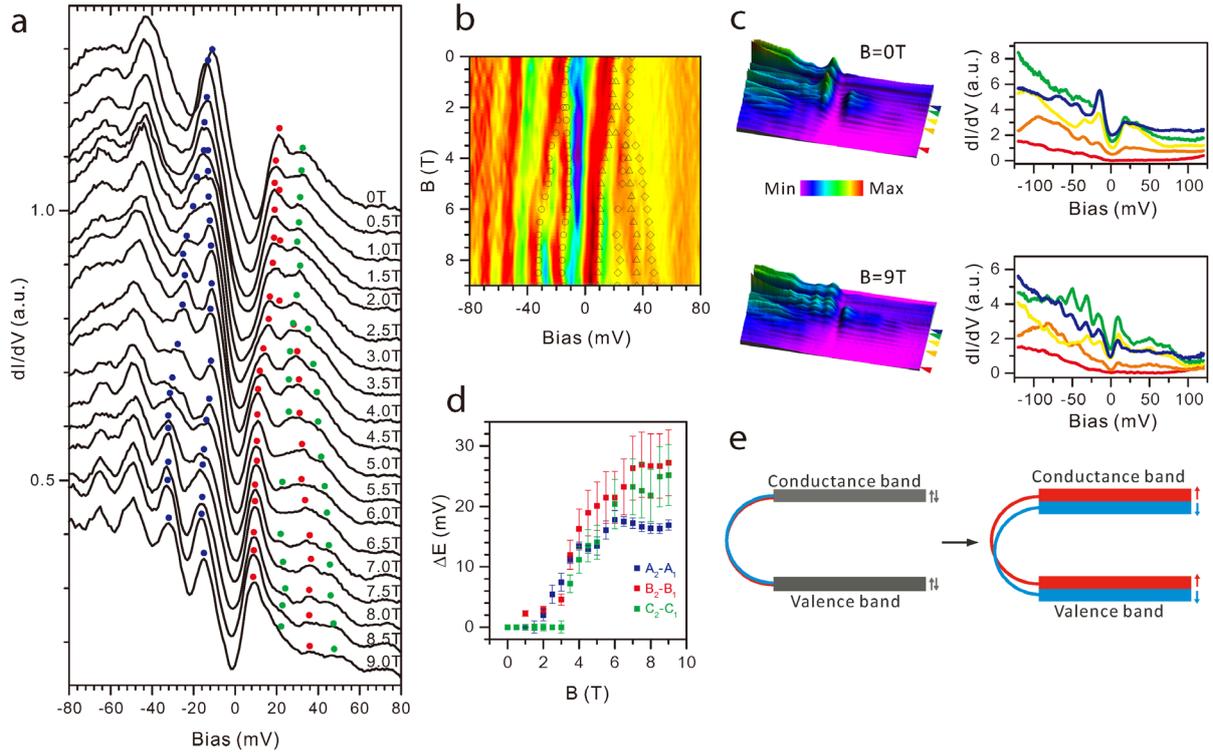

FIG. 3 (color online). (a) Differential conductance (dI/dV spectra) obtained at a certain position upon magnetic field, varying from 0 to 9 T, with an increment of 0.5 T. The three peaks in the band gap region are marked as A, B and C, and traced via colored dots, where A splits into $A_1$, $A_2$, the same for B and C. (b) The same spectra after differentiation is plotted in false colored mode. The peaks A, B and C are marked by open circle, triangle and diamonds. (c) dI/dV spectra (U = +250 mV, $I_t$ = 200 pA, $V_{osc}$ = 5 mV) taken across the step edge, under the magnetic field of 0 (up) and 9 Tesla (down) respectively. Five spectra are extracted and plotted in the right. The corresponding positions for taking the spectra are marked with colored triangles in the left. (d) Energy gap, defined as $\Delta E = E_2 - E_1$, versus B. (e) Sketch illustrating the close link between topological edge states splitting and the magnetic field induced bulk quantization. Left and right are sketches without and with an applied magnetic field, respectively.

# Supplementary Information:

## Robust topological edge states at the perfect surface step edge of topological insulator ZrTe$_5$

Xiang-Bing Li[1†], Wen-Kai Huang[1†], Yang-Yang Lv[2†], Kai-Wen Zhang[1], Chao-Long Yang[1], Bin-Bin Zhang[2], Y. B. Chen[1*], Shu-Hua Yao[2], Jian Zhou[2], Ming-Hui Lu[2], Li Sheng[1,3], Shao-Chun Li[1,3*], Jin-Feng Jia[3,4], Qi-Kun Xue[5], Yan-Feng Chen[2,3], Dingyu Xing[1,3]

[1] National Laboratory of Solid State Microstructures, School of Physics, Nanjing University, Nanjing, Jiangsu 210093, China

[2] National Laboratory of Solid State Microstructures, Department of Materials Science and Engineering, Nanjing University, Nanjing, Jiangsu 210093, China

[3] Collaborative Innovation Center of Advanced Microstructures, Nanjing University, Nanjing 210093, China

[4] Key Laboratory of Artificial Structures and Quantum Control (Ministry of Education), Department of Physics and Astronomy, Shanghai Jiao Tong University, Shanghai 200240, China

[5] State Key Laboratory of Low-Dimensional Quantum Physics, Department of Physics, Tsinghua University, Beijing 100084, China

†These authors contributed equally to this work.

*e-mail: ybchen@nju.du.cn; scli@nju.edu.cn.

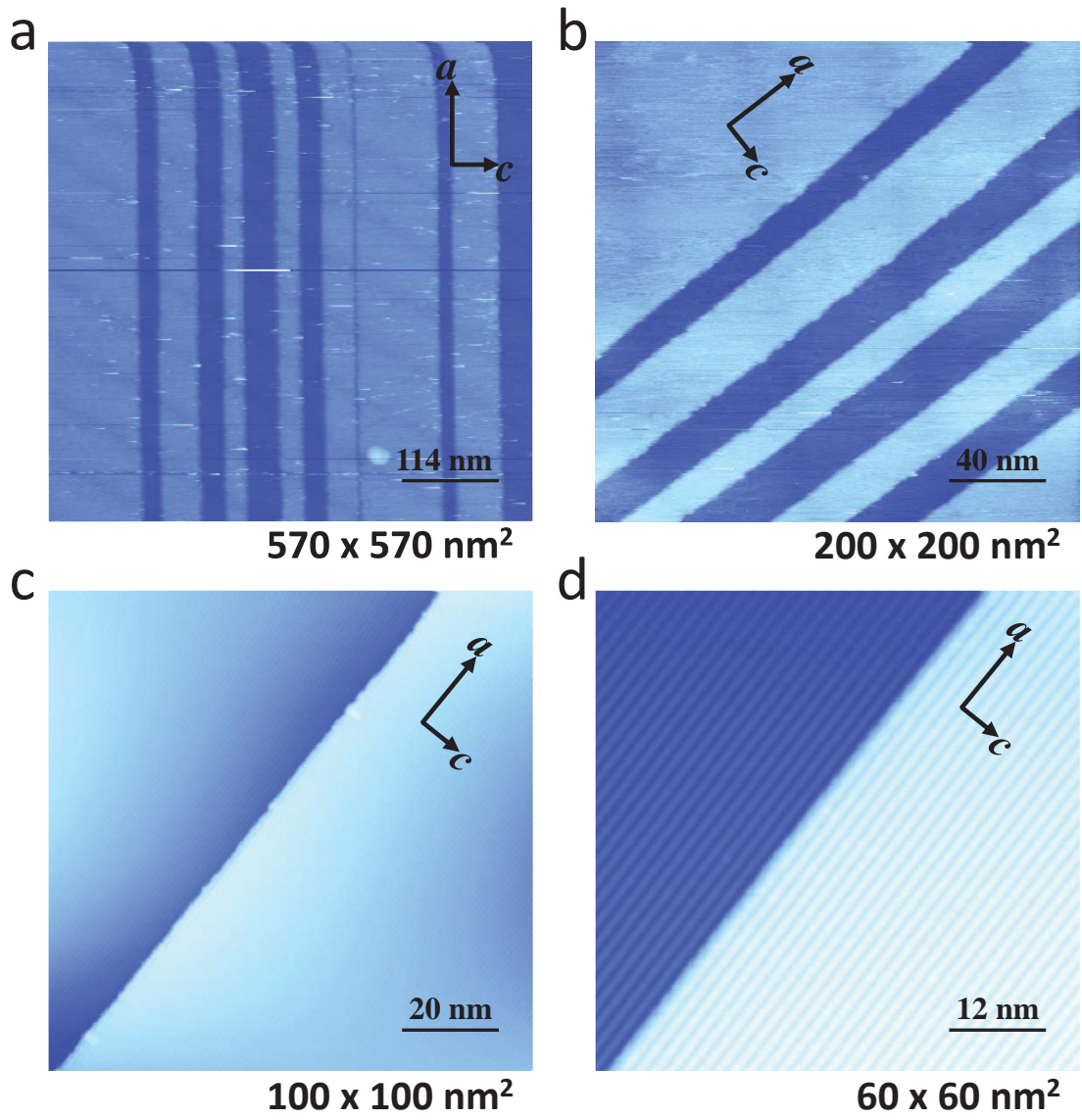

**Figure S1 Large scale STM images of ZrTe$_5$ surface. a, b** STM images of multiple long and straight surface steps along the a axis obtained at 80K (U = +1.0 V; I$_t$ = 100 pA; a, 570 × 570 nm²; b, 200 × 200 nm²); **c, d** STM image of single straight surface step along a axis obtain at 4K (U = +1.0 V; I$_t$ = 100 pA; c, 100 × 100 nm²; d, 60 × 60 nm²). Steps along the a axis were easier observed due to quasi 1D character, super straight in large scale, the height of single step is measured as ~ 0.8 nm.

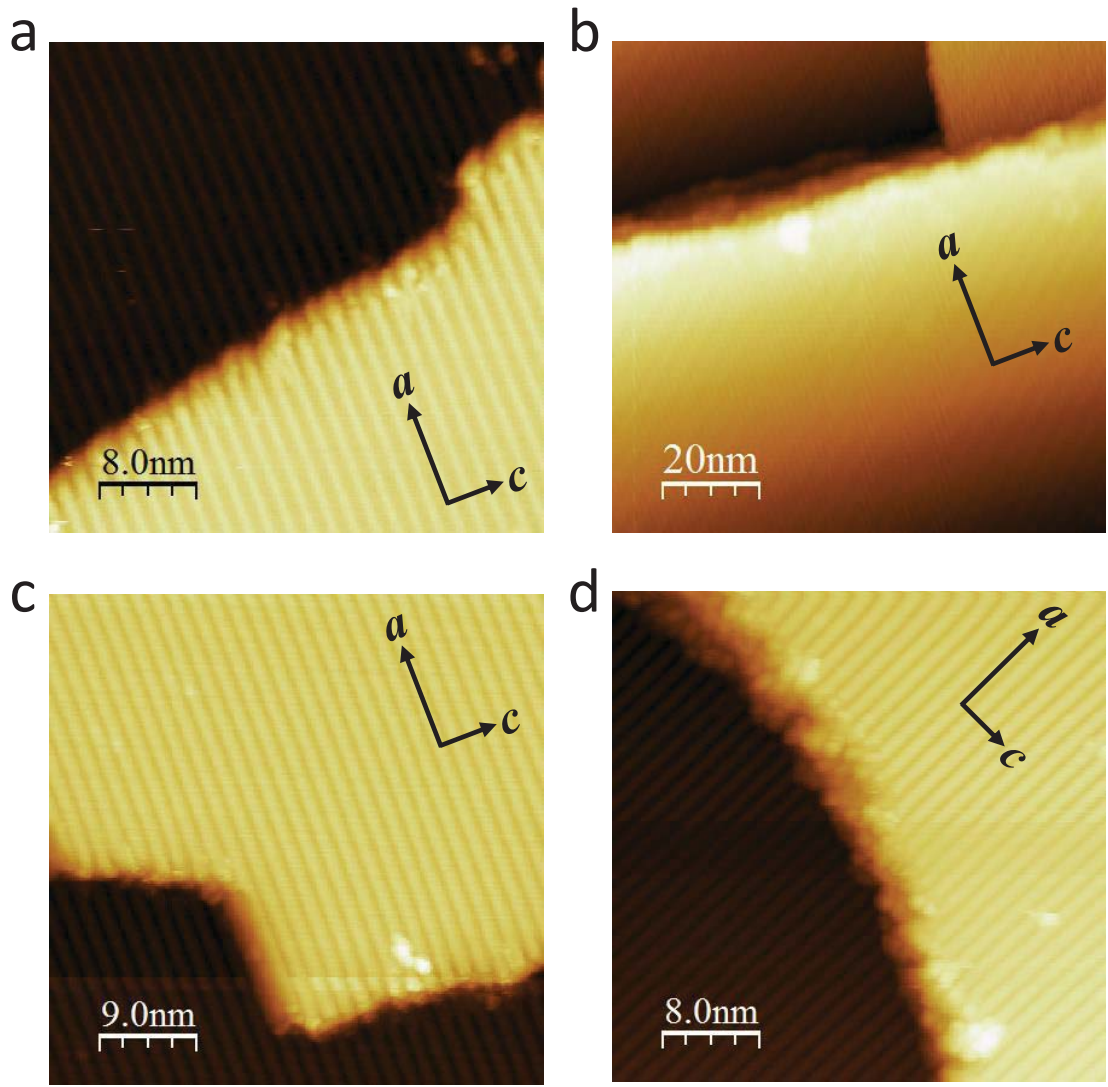

**Figure S2 Steps along the c axis at ZrTe$_5$ surface. a-d,** STM images of step along the c axis obtained at 4K (U = +1.0 V, I$_t$ = 100 pA). Such step can be seldom observed, and not straight, the height of single step is measured as ~ 0.8 nm.

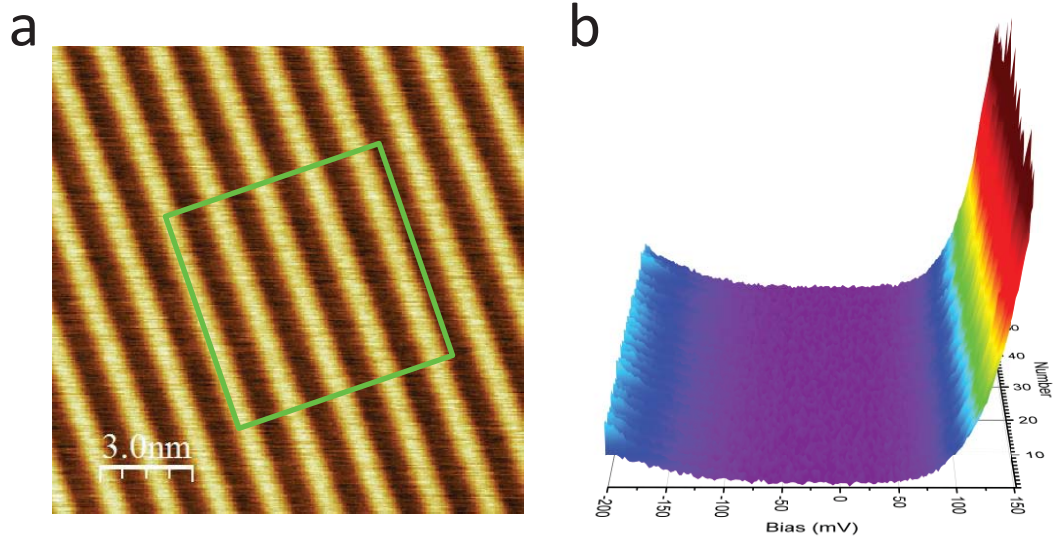

**Figure S3 The dI/dV spectra at various locations of ZrTe$_5$ terrace. a,** STM image of terrace away from the step obtained at 4K (U = +1.0 V, I$_t$ = 100 pA). **b,** dI/dV spectra taken at a 8 × 8 matrix in the green square shown in a (U = +250 mV, I$_t$ = 200 pA, V$_{osc}$ = 5 mV). These spectra is highly homogeneous due to high quality of ZrTe$_5$ single crystal, resemble to the dI/dV spectra in Fig. 1d.

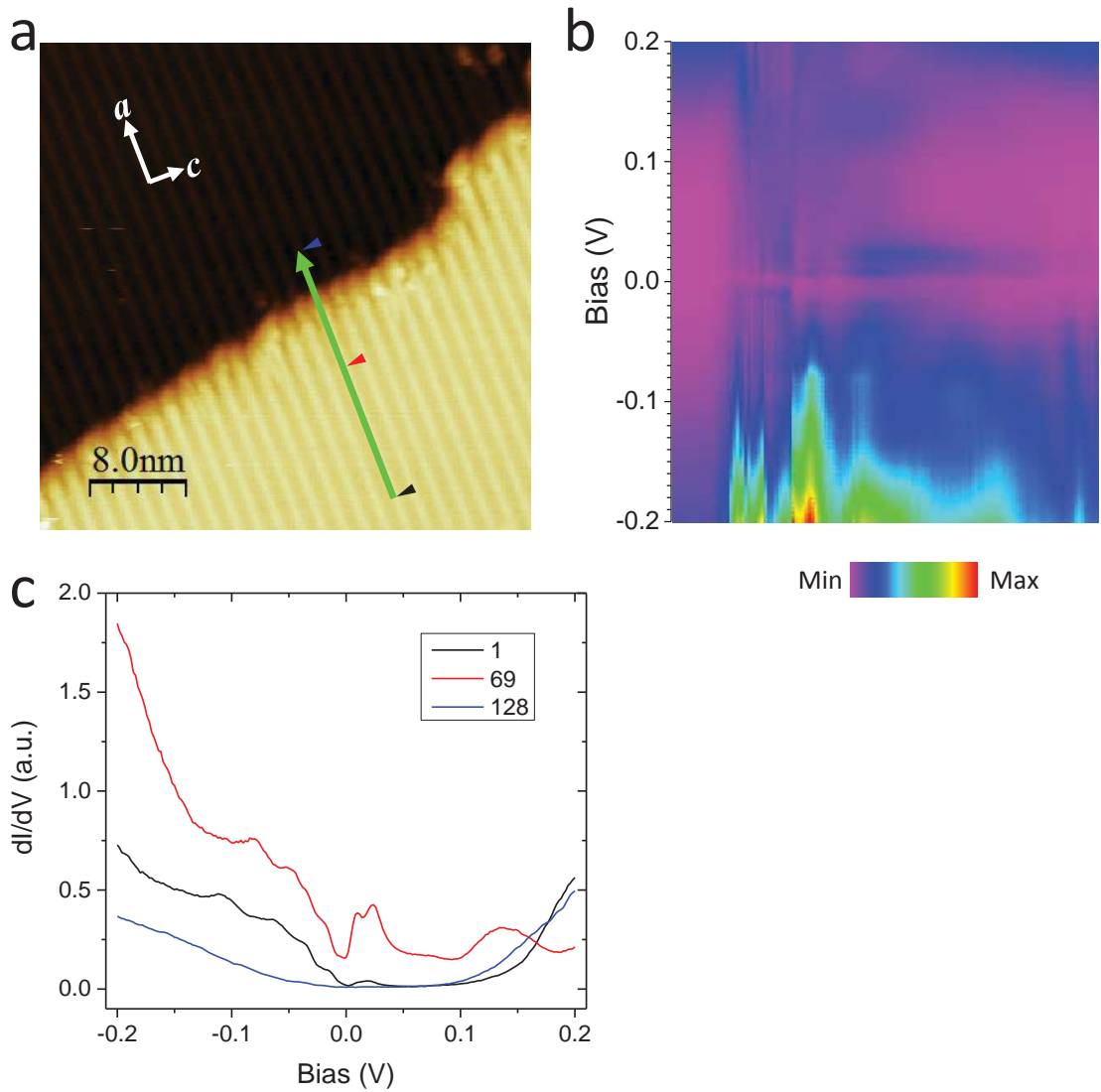

**Figure S4 The dI/dV spectra measurement taken across the step along the c axis. a,** STM image of the step along the c axis obtained at 4K (U = +1.0 V, $I_t$ = 100 pA). **b,** The dI/dV spectra measurement taken along the green line (20 nm in length) perpendicular to the step in a (U = +250 mV, $I_t$ = 200 pA, $V_{osc}$ = 5 mV). **c,** Three representative spectra extracted from b, the positions are marked in a as black, red and blue triangles.

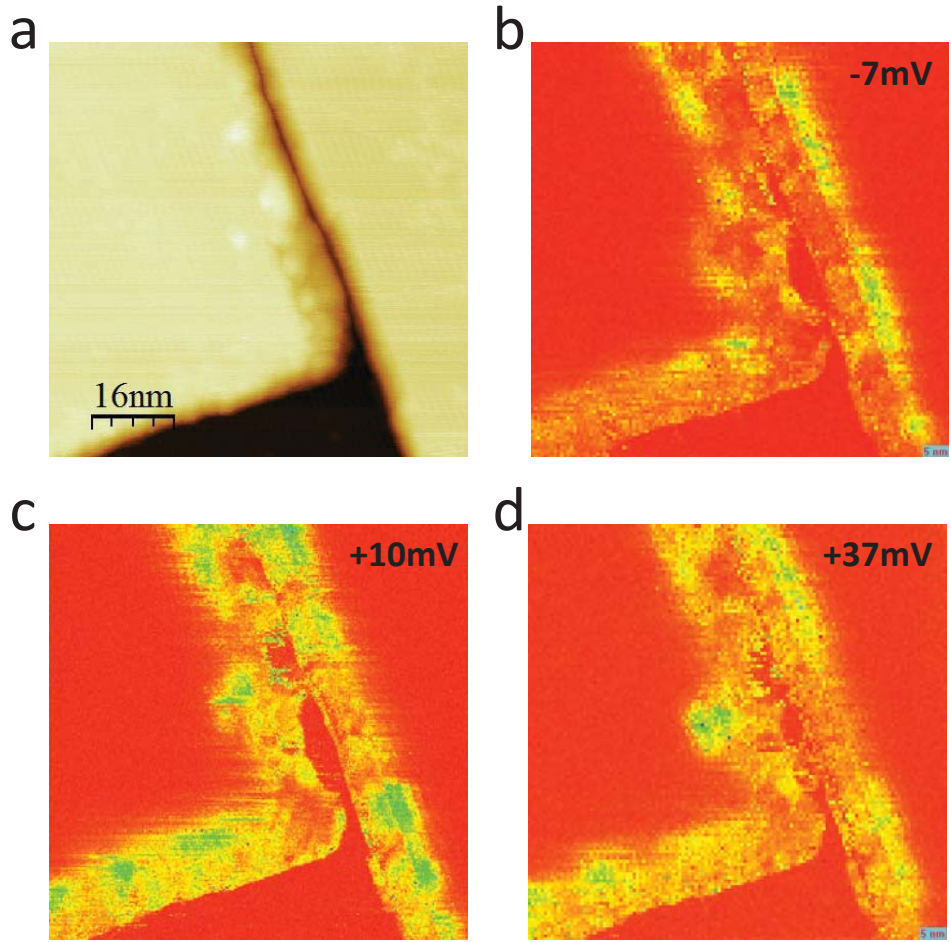

**Figure S5 The dI/dV mapping of different step. a,** STM image of surface step obtained at 4K (80 nm × 80 nm; U = +1.0 V, $I_t$ = 100 pA). **b-d,** The corresponding dI/dV mapping at three energies showing the topological edge state exist at all kinds of surface step edges, along the a and c axis (80 nm × 80 nm; U = +250 mV, $I_t$ = 200 pA, $V_{osc}$ = 5 mV).

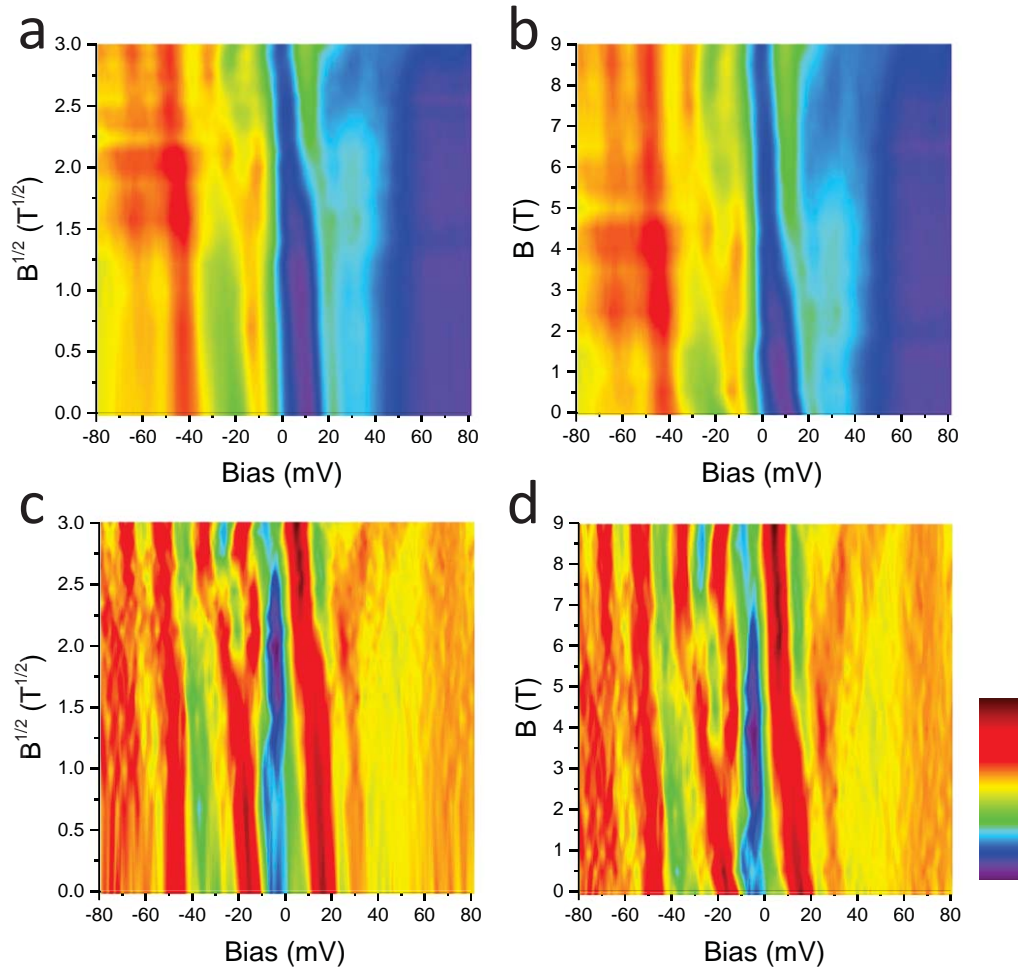

**Figure S6 Evolution of topological edge states in the presence of magnetic fields. a, b** The original dI/dV spectra obtained at certain position upon magnetic field at 4K, plotted in faulse colored mode. **c, d** The spectra after differentiating process. Left panel: dependence on sqrt(B); right panel: dependence on B, (U = +250 mV, $I_t$ = 200 pA, $V_{osc}$ = 5mV).

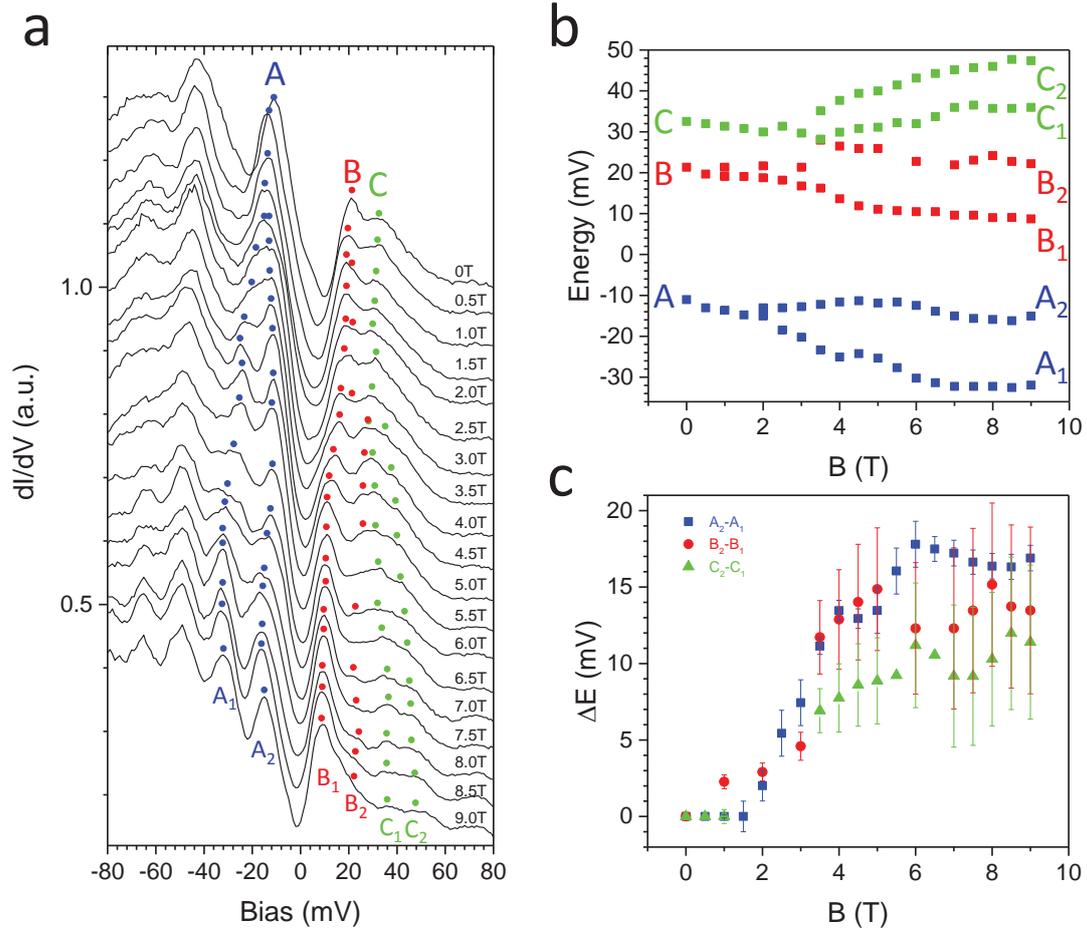

**Figure S7 An alternate assignment for peak A, B, and C. a,** The original dI/dV spectra obtained at certain position upon magnetic field at 4K, the splitting branches of peak B and C are marked differently from Fig. 3a. **b,** Energy of the splitting branches of peak A, B and C versus B. **c,** Energy gap, as defined $\Delta E = E_2 - E_1$ versus B.